\documentclass[12pt]{article}

\usepackage{amsfonts}
\usepackage{latexsym}

\begin{document}
\oddsidemargin=0.8cm

\def\bib{\bibitem}
\def\be{\begin{equation}}
\def\ee{\end{equation}}
\def\ba{\begin{eqnarray}}
\def\eea{\end{eqnarray}}
\def\df{\stackrel{\rm def}{=} }

\begin{center}
{\large \bf A Theory of the Casimir Effect for Compact Regions}
\end{center}
\vspace{0.2cm}
\begin{center}
Luiz A. Manzoni\footnote{E-mail: lmanzoni@fma.if.usp.br}  and  Walter F. Wreszinski\footnote{E-mail: wreszins@fma.if.usp.br}\linebreak[1]
\linebreak[1]
Instituto de F\'{\i}sica\linebreak[1]
Universidade de S\~ao Paulo\linebreak[1]
Caixa Postal 66318 \linebreak[1]
05315-970 - S\~ao Paulo, S.P. \linebreak[1]
Brazil\linebreak[1]
\linebreak[1]

\end{center}

\begin{abstract}
{\footnotesize
We develop a mathematically precise framework for the Casimir effect. Our working hypothesis, verified in the case of parallel plates, is that only the regulariza{\-}tion-independent Ramanujan sum of a given asymptotic series contributes to the Casimir pressure. As an illustration, we treat two cases: parallel plates, identifying a previous cutoff-free version (by G. Scharf and W.W.) as a special case, and the sphere. We finally discuss the open problem of the Casimir force for the cube. We propose an Ansatz for the exterior force and argue why it may provide the exact solution, as well as an explanation of the repulsive sign of the force.
}
\end{abstract}



\section{\large A General Framework for the Casimir Effect}

Significant progress on the Casimir effect from the experimental point of view occurred in recent times \cite{LMR}. In spite of that, several theoretical problems remain, such as a real explanation of the sign of the force in the case of compact regions. The situation is worse with regard to a mathematically precise framework for the effect, due to the cutoff (or regularization) dependence of the energy, a fact emphasized by C. R. Hagen in \cite{Hag01} and somewhat less emphatically by P. Candelas in \cite{Can82}. The physical reason why divergences occur is well understood \cite{DCa79} and is that the boundaries are treated by quantizing the radiation field with mode functions  \cite{Mil82} which are adapted to the type of (classical) boundary conditions (b.c.), e.g., Dirichlet or Neumann. However, real boundaries consist of electrons and ions and such b.c. are not justified except if the particles act collectively in an essentially classical manner \cite {Mil82}, which is a priori not the case \cite{DCa79}, and our ignorance in dealing with this fact is signalled by divergences.

Divergences are, of course, well-known in field theory, but they arise here in a different way, as explained above. Mathematical physicists, and several theoretical physicists, agree that a mathematically precise framework to cope with these divergences would be conceptually useful. Such frameworks exist in field theory (see \cite{Hep69, Sch95} and references given there). A cutoff-free or ``finite" theory of the Casimir effect (in the spirit of \cite{Sch95}) was attempted by G. Scharf and one of us (W.W.) in \cite{SWr92}. It requires, however, the use of periodic b.c., which are unphysical in the case of the electromagnetic field.

In this paper we present a thorough derivation of the results first announced in \cite{MWr01}. We reconsider the problem introducing {\it ab initio} an ultraviolet cutoff ($1/\Lambda$). The Casimir energy (CE) $E_{\rm vac}(\Lambda )$ would diverge if the limit $\Lambda \rightarrow 0$ was taken, but we do not need to do so, because the Casimir {\it pressure} depends only on the $\Lambda$-independent term in the asymptotic expansion, which is the (RI) cutoff independent term of the Ramanujam sum of a divergent series. This idea is due to K. Dietz \cite{Die88}. In section \ref{plates} we show how the result of \cite{SWr92} for the parallel plates is recovered as a special case. Some of the ideas of \cite{SWr92} are also used and summarized below, for convenience.

Following \cite{Boy68}, consider an electromagnetic field at $T=0$ enclosed in cavities of identical shape, but made of different materials, the latter providing natural cutoffs for the high-frequency spectrum of zero point modes. The vacuum energy is thus given by
\be
E_{\rm vac}=\frac{\hbar}{2}\sum_{\alpha} \omega_{\alpha}C_{\alpha}(\Lambda )\; ,
\label{ereg}
\ee

\noindent
with $C_{\alpha}(\Lambda )$ material dependent cutoff functions dependent on a variable $\Lambda$ with dimensions of length, which we normalize by
\be
C_{\alpha}(\Lambda )|_{\Lambda =0}=1\; .
\label{normaliz}
\ee

\noindent
Since $E_{\rm vac}$ has dimension $(length)^{-1}$ in natural units, it may be written as an (asymptotic) series
\be
E_{\rm vac} = a_0 L^3\Lambda^{-4}+a_1L^2 \Lambda^{-3} + a_2L \Lambda^{-2}+a_3 \Lambda^{-1}+a_4L^{-1}+a_5L^{-2} \Lambda + \ldots \; ,
\label{easym}
\ee

\noindent
where $L$ is a length characterizing the spatial extension of the cavity. Dietz conjectured \cite{Die88} that by a theorem of Ramanujan the  $\Lambda$-independent term $a_4L^{-1}$ in (\ref{easym}) is {\it independent of the regularization} (i.e., of the set $\{ C_{\alpha}(\Lambda )\}$ in (\ref{ereg})) provided (\ref{normaliz}) holds. We shall return to this conjecture later.

In this paper we consider as in \cite{SWr92} the prototypical example of massless scalar field confined in a compact region $K$ (a compact manifold with boundary) -- the modifications introduced by considering the full electromagnetic field will be mentioned later. We show that the $\omega_{\alpha}$ in (\ref{ereg}) should be identified with the eigenvalues of the {\it square root} of the Laplace-Beltrami operator. This is not unexpected, because the relativistic energy is $| \vec{k}|=({\vec{k}}^2)^{1/2}$, but it has important consequences for expansion (\ref{easym}). Consider \cite{SWr92} the field $A(x)$ quantized in infinite space
\be
A(x)=\frac{1}{(2\pi )^{\frac{3}{2}}}\int \frac{d^3k}{\sqrt{2\omega}}\left[ a(\vec{k})e^{-ik\cdot x}+ a^+(\vec{k})e^{ik\cdot x}\right]\; ;
\ee
\be
\left[A_-(x), A_+(y)\right]=\frac{1}{i}D_0^{(+)}(x-y)\; ;
\ee
\be
D_0^{(+)}(x)= \frac{i}{(2\pi )^3}\int \frac{d^3k}{2 |\vec{k}|}e^{-ik\cdot x} =-\frac{i}{4\pi^2}\frac{1}{(x_0-i0)^2-\vec{x}^2} \; .
\ee

Time evolution is generated by the Hamiltonian $H=\int d^3xH(x)$, whose density can be written in the form
\be
H(x)=\frac{1}{2}:\left[\left(\frac{\partial A}{\partial x_0}\right)^2-A\frac{\partial^2 A}{\partial x_0^2}\right]:\; .
\ee

\noindent
Normal ordering is defined in momentum space. In order to go over to a geometry with {\it boundaries}, we should formulate it in $x$-space by the point-splitting technique:
\ba
:\left(\frac{\partial A}{\partial x_0}\right)^2: &=& \lim_{y\rightarrow x} :\frac{\partial A(x)}{\partial x_0}\frac{\partial A(y)}{\partial y_0}: \nonumber \\  \\
&=&\lim_{y\rightarrow x} \left\{ \frac{\partial A(x)}{\partial x_0}\frac{\partial A(y)}{\partial y_0} +\frac{1}{i}\frac{\partial^2}{\partial x_0^2}D_0^{(+)}(x-y)\right\}\; . \nonumber 
\eea

\noindent
Finally,
\be
H(x)= \lim_{y\rightarrow x} \left\{ \frac{1}{2}\frac{\partial A(x)}{\partial x_0}\frac{\partial A(y)}{\partial y_0} -\frac{1}{2} A(x)\frac{\partial^2 A(y)}{\partial y_0^2} +\frac{1}{i}\frac{\partial^2}{\partial x_0^2}D_0^{(+)}(x-y)\right\}\; .
\label{hamilt}
\ee

\noindent
Taking into account that real boundaries consist of electrons and ions and the field which interacts with them is quantized in {\it infinite space}, we consider (\ref{hamilt}) to be the Hamiltonian density describing the field both free and with boundaries. In the latter case, however, the first two terms in (\ref{hamilt}) must be defined in the adequate Fock space, i.e., the concrete representation of the field operator is dictated by the geometry. Consider a compact region $K$ and Dirichlet b.c. $A(x)=0$ for $\vec{x} \in \partial K$. Then $A(x)$ may be expanded as follows:
\be
A(x)=\sum_n\frac{1}{\sqrt{2\omega_n}}\left[ a_nu_n(\vec{x})e^{-i\omega_n x_0}+ a^+_nu_n(\vec{x})e^{i\omega_n x_0}\right]\; ,
\ee

\noindent
where $u_n$ are normalized real eigenfunctions of the Laplacian in $K$, satisfying Dirichlet or Neumann b.c. (discrete spectrum):
\be
-\Delta u_n(\vec{x})=\omega^2_nu_n(\vec{x})\; .
\label{lapla}
\ee

\noindent
The concrete Fock representation is now specified by regarding $a^+_n$, $a_n$ as emission and absorption operators ($[a_n,a^+_m]=\delta_{nm}$) and defining the vacuum by
\be
a_n\Omega =0 \;\; \forall \;n\; .
\label{aniq}
\ee

\noindent
We thus find in this Fock representation:
{
\newcounter{let}
\setcounter{let}{1}
\renewcommand{\theequation}{\arabic{equation}\alph{let}}
\ba
H(x)&=&\frac{1}{2};\left(\frac{\partial}{\partial x_0}A(x)\right)^2;-\frac{1}{2};A(x)\frac{\partial^2}{\partial x_0^2}A(x); \nonumber \\  \label{hmom} \\ \addtocounter{let}{1}\addtocounter{equation}{-1}
&+&\frac{1}{i}\lim_{y\rightarrow x}\frac{\partial^2}{\partial x_0^2}\left\{D_0^{(+)}(x-y)-D_K^{(+)}(x,y)\right\} \; ,\nonumber
\eea

\noindent
where
\be
D_K^{(+)}(x_0-y_0,\vec{x},\vec{y})=i\sum_n\frac{1}{2\omega_n}u_n(\vec{x})u_n(\vec{y})e^{-i\omega_n (x_0-y_0)} \; ,
\label{d+k}
\ee
}

\noindent
and the semicolons in (\ref{hmom}) denote normal ordering with respect to the new emission and absorption operators $a^+_n$ and $a_n$. Notice that $D_0^{(+)}$ is the solution of the wave equation $\Box D_0^{(+)}=0$ with initial conditions
{
\setcounter{let}{1}
\renewcommand{\theequation}{\arabic{equation}\alph{let}}
\ba
D_0^{(+)}(+0,\vec{x})&=& \frac{i}{4\pi^2}\frac{1}{\vec{x}^2+i0}\; ;\\ \nonumber \\ \addtocounter{let}{1}\addtocounter{equation}{-1}
\left(\partial_0D_0^{(+)}\right)(0,\vec{x})&=&\frac{1}{2}\delta (\vec{x})\; ,
\eea
}

\noindent
$D_0^{(+)}(+0,\vec{x})$ is the Green's function of the square root of $-\Delta$ in infinite space \cite{SWr92}. Similarly
{
\setcounter{let}{1}
\renewcommand{\theequation}{\arabic{equation}\alph{let}}
\ba
\left. \left(\partial_0D_K^{(+)}\right)(x,y)\right|_{y_0=x_0}&=&\frac{1}{2}\delta (\vec{x}-\vec{y})\; ;  \\ \nonumber \\ \addtocounter{let}{1} \addtocounter{equation}{-1}
D_K^{(+)}(+0,\vec{x},\vec{y})&=&\frac{i}{2}(-\Delta_K)^{-\frac{1}{2}}(\vec{x}-\vec{y})\; , 
\eea
}

\noindent
where $\Delta_K$ denotes the Laplacian on $K$, with Dirichlet or Neumann b.c..

We now consider two types of cutoff functions, one of them general, satisfying (\ref{normaliz}), the other special, of type
{
\setcounter{let}{1}
\renewcommand{\theequation}{\arabic{equation}\alph{let}}
\be
C_{\alpha}(\Lambda )=C(\Lambda \omega_{\alpha} )\; , 
\label{esp}
\ee
\addtocounter{let}{1}\addtocounter{equation}{-1}

\noindent
satisfying
\be
C(0)=1\; .
\label{noresp}
\ee
}

\noindent
We shall also interested in a particular case of (\ref{esp}), namely
\be
C(k)=e^{-\Lambda k}\;\;\;\;\; k\geq 0\; .
\label{ck}
\ee

In terms of the special choice (\ref{ck}), we may, by (\ref{hamilt}), (\ref{aniq})-(\ref{d+k}), compute a regularized vacuum energy density $H_{\rm vac}(x, \Lambda)$ in the following way:
\ba
H_{\rm vac}(x, \Lambda)&=&\frac{1}{2}\frac{\partial}{\partial \Lambda}\left\{\frac{1}{(2\pi )^3}\int d^3k \left.e^{-i[k_0\tau -\vec{k}\cdot(\vec{x}-\vec{y})]}\right|_{ \stackrel{\scriptscriptstyle \vec{y}=\vec{x} }{\scriptscriptstyle \tau =0} }
C( k_0)\right. \nonumber \\ \label{h} \\
&-&\left. \sum_n\left[ u_n(\vec{x})\right]^2C(\omega_n )\right\}\; .\nonumber
\eea

\noindent
As an aside, notice that (\ref{ck}) corresponds to ascribe a small imaginary part $-i\Lambda$ to $x_0-y_0=\tau$, and thus represents a ``natural'' choice, akin to the principal value in distribution theory \cite{GSh75}. For this special case (\ref{ck}),
{
\setcounter{let}{1}
\renewcommand{\theequation}{\arabic{equation}\alph{let}}
\be
H_{\rm vac}(\vec{x}, \Lambda )=\frac{1}{2}\frac{\partial}{\partial \Lambda}\left[ P(\vec{x},\vec{x};\Lambda )-P_0(\vec{x},\vec{x};\Lambda )\right]\; ,
\label{hp}
\ee
\addtocounter{let}{1}\addtocounter{equation}{-1}
\noindent
where $P$, $P_0$ satisfy the ``heat equation''
\be
\left(\frac{\partial}{\partial\Lambda}-(-\Delta_{\vec{x}})^{\frac{1}{2}}\right)P(\vec{x},\vec{y};\Lambda )=0\; ,
\label{heat}
\ee
\addtocounter{let}{1}\addtocounter{equation}{-1}
\noindent
with the b.c.
\be
P(\vec{x},\vec{y}; \Lambda )=0\hspace{0.5cm}{\rm if}\; \vec{x}\;\; {\rm or}\; \vec{y}\; \in \; \partial K\; ,
\ee
}
\noindent
in the case of Dirichlet b.c..

There exist methods to compute the asymptotic expansion (in $\Lambda$) of the quantity in brackets in (\ref{hp}) \cite{Kan77}, which solve the problem in principle, but the actual form (\ref{easym}), with the given coefficients, depends on the details of the discrete (eigenvalue) spectrum of $(-\Delta)^{\frac{1}{2}}$.

Let now $L$ be a linear dimension of the compact region $K\equiv K_L$ and $M$ a linear dimension of a region $K_M$ of which $K_L$ is a subset. Typically, if $K_L$ is a cube of side $L$, $K_M$ is a cube of side $M>L$ concentric with $K_L$, and similarly for a sphere or other manifolds. It is correct to impose the same b.c. (e.g. Dirichlet or Neumann) on $K_M$ in order to define the outer Casimir problem \cite{BHa98, Boy68}. In fact, previous work on the sphere using the Sommerfeld radiation condition was not correct, although the results were right, because it did not lead to real eigenvalues \cite{NPi98}. Define
\be
E_{\rm vac}(L,\Lambda ,M)=E_{\rm vac}^{\rm inner}(L,\Lambda )+E_{\rm vac}^{\rm outer}(L,\Lambda ,M)\; ,
\label{ellm}
\ee

\noindent
where
\be
E_{\rm vac}^{\rm inner}(L, \Lambda ) \equiv \int_{K_L}d^3x\; H(\vec{x},\Lambda )\; ,
\label{ellmin}
\ee

\noindent
and
\be
E_{\rm vac}^{\rm outer}(L, \Lambda , M) =\int_{K_M\backslash K_L}d^3x\; \tilde{H}(\vec{x},\Lambda )\; .
\label{ellmou}
\ee

\noindent
As previously remarked, if Dirichlet b.c. are imposed on $K_L$, $H$ (respec. $\tilde{H}$) is the density (\ref{h}) with the $\{ u_n \}$ defined by Dirichlet b.c. imposed on $K_L$ (respec. $K_L$ and $K_M$). Analogous definitions hold for other b.c. (e.g. Neumann or mixed). If (\ref{ereg}), (\ref{normaliz}) is adopted, the second sum in (\ref{ellm}) refers, then, to the modes $\omega_n$ corresponding to the solution of (\ref{lapla}) in $K_M\backslash K_L$, with the above-mentioned b.c.. Suppose that both $E_{\rm vac}^{\rm inner}(L, \Lambda )$ and $E_{\rm vac}^{\rm outer}(L, \Lambda , M)$ have asymptotic series (\ref{easym}), and let $E_{\rm vac}^{\rm inner}(L)(\equiv a_4^{\rm inner}/L)$ and $E_{\rm vac}^{\rm outer}(L, M)$ be the corresponding $\Lambda$-independent terms. Then the Casimir pressure on the boundary surface $p_C(L)$ (a measurable quantity) is defined by the thermodynamic formulae (zero absolute temperature):
{
\setcounter{let}{1}
\renewcommand{\theequation}{\arabic{equation}\alph{let}}
\be
p_C(L)=p_C^{\rm inner}(L)-p_C^{\rm outer}(L)\; ,
\label{p}
\ee
\addtocounter{let}{1}\addtocounter{equation}{-1}

\noindent
where the relative minus sign takes into account that $p_C^{\rm outer}$ refers to a normal vector pointing inwards towards $K_L$, while $p_C^{\rm inner}$ refers to a normal vector pointing outwards, and
\begin{eqnarray}
&&p_C^{\rm inner}(L)=-\frac{\partial E_{\rm vac}^{\rm inner}(L)}{\partial V_{\rm inner}(L)}\; ; \label{pin} \\ \nonumber 
\addtocounter{let}{1}\addtocounter{equation}{-1} \\
&&p_C^{\rm outer}(L)=-\lim_{M\rightarrow \infty}\frac{\partial E_{\rm vac}^{\rm outer}(L, M)}{\partial V_{\rm outer}(L, M)}\; , \label{pou}
\end{eqnarray}
}

\noindent
and an important feature of the {\it thermodynamic limit} \cite{Rue} is that the derivative in (\ref{pou}) is taken with $M$ {\it fixed}, only $L$ varies. 

It is essential that the CE be independent of the cutoff function $C$ in (\ref{ereg}) or (\ref{esp}) provided it satisfies (\ref{normaliz}) or (\ref{noresp}). As remarked in \cite{Die88}, a necessary condition for this regularization independence (RI) to hold is that (\ref{easym}) contain no logarithmic terms, because, otherwise, the ``$\Lambda$-independent term'' is obviously ill-defined. For the cube there are no such terms in (\ref{easym}), but such is not the case for the sphere; however such terms may be omitted in the case of the sphere because they cancel in the expression $E_{\rm vac}^{\rm inner}(L,\Lambda )+ E_{\rm vac}^{\rm outer}(L, \Lambda , M) $ which have an asymptotic series (\ref{easym}) as $M\rightarrow \infty$, so that for the sphere of radius $a$ we have:
\be
p_C(a)=-\frac{1}{4\pi a^2}\frac{\partial}{\partial a} \frac{a_4^{\rm sphere}}{a}= \frac{a_4^{\rm sphere}}{4\pi a^4}\; .
\label{psin}
\ee

\noindent
A full proof of RI is given in section 2 for parallel plates. We shall leave a more detailed discussion of higher dimensional cases \cite{Ros64} to a further publication, but we wish to make a few important remarks:
\begin{itemize}

\item[$a$)] the $\Lambda$-independent term in (\ref{easym}) should coincide with the $\Lambda$-independent term of the Ramanujan sum of a divergent series of positive terms, such as (\ref{ereg}), with $C_{\alpha}(\Lambda )\equiv 1$ see \cite{Har49} (p. 318 ff.) and section 1. According to this concept, for instance
\begin{eqnarray*}
&&1+1+\cdots +1+\cdots = -\frac{1}{2}\;(\Re ,0)\; ; \\ \\
&&1+2+3+\cdots =-\frac{1}{12}\; (\Re,0)\; ,
\end{eqnarray*}

\noindent
taking the origin as reference point (see \cite{Har49}, 13.10.11). This is proved in section \ref{plates} for parallel plates, and is the basis of RI;

\item[$b$)] the present definition of the CE is mathematically rigorous. In particular, the limit $\Lambda \rightarrow 0$ is never taken. In fact, (\ref{easym}) shows that, in general, it does not exist (an exception is the Casimir effect for parallel plates with periodic b.c., see \cite{SWr92} and section \ref{plates}). The reason for this is that we do not know how to treat the surface properly in microscopic terms, a formidable problem (see the conclusion);

\item[$c$)] RI justifies the definition of the $\Lambda$-independent term in (\ref{easym}) as the CE physically: it reflects the field theoretic structure of the vacuum state which is independent of the cavity materials \cite{Die88}. It is also expected to be the only term in (\ref{easym}) which contributes to the {\it pressure}: this was proved in \cite{Die88} for parallel plates.

\end{itemize}

Section \ref{plates} entails a complete proof of RI for parallel plates (a ``limit'' of a compact region), as well as the explanation of the ``theory without cutoffs'' for the case of periodic boundary conditions in ref. \cite{SWr92}. The sphere is also treated a an illustration, in section \ref{sphere}. In contrast to the parallel plates, the force for the sphere is repulsive, as known since the pioneering work of Boyer \cite{Boy68}. The nature of our derivation, which is a adaptation of the ideas and results of \cite{NPi98} and \cite{BHa98} to our framework, does not convey an intuitive ``explanation'' of the sign of the force. This is a difficult problem because the CE is a sum of {\it fluctuations} of the electric and magnetic fiels in the vacuum state. A basic issue is: if the flat parallel plate geometry is changed to a compact manifold with boundary, how does the sign of the force change and why? This question is most clearly analysed in the case of the cube, which is the simplest deformation of the parallel plates geometry. In section \ref{cinner} we consider the {\it interior} problem for the cube, using the method of the Poisson summation formula used in \cite{SWr92}. This method had already been used for the same purpose in \cite{RZV77}. Since this reference is not readily available, we include our (independent) derivation which generalizes \cite{RZV77} in the sense that we obtain the full asymptotic formula, and fits nicely into the present framework. It should be also remarked that it coincides with the numerical result of \cite{AWo82}. The inner problem leads, however, to an {\it attractive} force, while the result for the sphere leads us to expect a repulsive force. Therefore, the repulsive nature must be due entirely to the {\it exterior} pressure.

In section \ref{couter} we introduce an Ansatz to solve the exterior problem for the cube, which leads to a repulsive force. We also have applied the Ansatz to the only known soluble case with flat geometry, i.e., the case of parallel plates (Appendix A). From the analysis of this soluble case we identify the physical reason why our Ansatz does not modify the pressure: the Ansatz introduces some extra stresses, but these are {\it parallel} to the plates (faces of the cube) so that the pressure {\it on} the plates (faces) is insensitive to these additional stresses, then providing the correct result for the plates and (we believe) for the cube. If the latter conjecture is true, an ``explanation'' of the sign of the force also follows. This is left to the conclusion and open problems in section \ref{conclusion}.

We have used a very general class of cutoffs in {\it momentum} space for parallel plates. For the cube the proof is essentially the same as for parallel plates, but there are some subtleties in the case of the sphere which have not yet been fully worked out. Nevertheless, it is an {\it open problem} whether only the cutoff-independent part of the CE is relevant to the {\it pressure}, except in the explicit case of parallel plates \cite{Die88}. We shall admit this as a working hypothesis throughout.

\section{\large Parallel Plates}
\label{plates}

We consider the problem of parallel plates, with distance $d$ along the $z$-axis; take the positions of the plates at $z=0$ and $z=d$, and adopt the form (\ref{esp}) in (\ref{h}), (\ref{ellm}) with Dirichlet b.c. (Neumann b.c. yield the same results). The {\it inner} Casimir problem corresponds to the region $K_L=K_d=\{\vec{x} \in \mathbb{R}^2 \times [0,d]\}$, and the {\it outer} one to the region $K_R\backslash K_L =\{\vec{x} \in \mathbb{R}^2 \times [d,d+R]\} \cup \{\vec{x} \in \mathbb{R}^2 \times [-R,0]\}$. The eigenfunctions associated to the inner problem are
\be
u_n^{\rm inner}(k_x,k_y)=\frac{1}{2\pi}\sqrt{\frac{2}{d}}\sin\left(\frac{n\pi}{d}z\right)e^{i(k_xx+k_yy)}\;\;\;\;\;n=1,2,3\cdots\; , 
\ee

\noindent
corresponding to the eigenvalues of $(-\Delta)^{\frac{1}{2}}$ given by
\be
\omega^{\rm inner}_{n,k_x,k_y}=\sqrt{\left(\frac{n\pi}{d}\right)^2+k_x^2+k_y^2}\; ,
\ee

\noindent
in (\ref{lapla}). The outer eigenfunctions are
\ba
&&u_n^{\rm outer, 1}(k_x,k_y)=\frac{1}{2\pi}\sqrt{\frac{2}{R}}\sin\left(\frac{n\pi}{R}(z-d)\right)e^{i(k_xx+k_yy)}\; ;\nonumber \\ \label{eigou}\\
&&u_n^{\rm outer, 2}(k_x,k_y)=\frac{1}{2\pi}\sqrt{\frac{2}{R}}\sin\left(\frac{n\pi}{R}z\right)e^{i(k_xx+k_yy)}\; ,\nonumber
\eea

\noindent
with eigenvalues
\be
\omega^{\rm outer}_{n,k_x,k_y}=\sqrt{\left(\frac{n\pi}{R}\right)^2+k_x^2+k_y^2}\; .
\label{omou}
\ee

We first adopt the choice (\ref{ck}). Introducing polar coordinates in the $x$-$y$ plane, we calculate the first (inner) sum in (\ref{ellm}) (we do {\it not} integrate along $(x,y)\in \mathbb{R}^2$, which would yield $+\infty$). The proper way to do this is to limit the $(x$-$y)$-plane integration to a finite region with area $A$, and then take the limit for ${\cal E}=\frac{E}{A}$ (this procedure yields the same results presented here and we omit it for brevity):

\ba
{\cal E}_{\rm vac}^{\rm inner}(\Lambda ,d)&=&\frac{1}{2(2\pi)^2}\left\{-2d\int_0^{\infty}dk \;k^3\;e^{-\Lambda k}\right. \nonumber \\ \label{ein} \\
&+&\left. 2\pi\sum_{n=1}^{\infty}\int_0^{\infty}dk\; k \;e^{-\Lambda \sqrt{\left(\frac{n\pi}{d}\right)^2+k^2}}\;\sqrt{\left(\frac{n\pi}{d}\right)^2+k^2}\right\} \; . \nonumber
\eea

Performing the change of variable $k_n'= \sqrt{\left(\frac{n\pi}{d}\right)^2+k^2}$ in the second integral in the r.h.s. of (\ref{ein}) we obtain
\ba
{\cal E}_{\rm vac}^{\rm inner}(\Lambda ,d)&=&\frac{1}{2(2\pi)^2}\left\{-2d\int_0^{\infty}dk \;k^3\;e^{-\Lambda k}+ 2\pi\sum_{n=1}^{\infty}\int_{\frac{n\pi}{d}}^{\infty}dk_n'\; k_n'^2 \;e^{-\Lambda k_n'} \right\} \nonumber \\ \nonumber \\
&=&\frac{d}{(2\pi)^2}\left\{ -6\Lambda^{-4}+ \frac{\partial^2}{\partial\Lambda^2}\left[ \frac{1}{\Lambda^2}\frac{\frac{\Lambda\pi}{d}}{e^{\frac{\Lambda\pi}{d}}-1}\right]\right\} \; , \label{einc}
\eea

\noindent
we now use the expansion (\cite{Har49}, p. 320) in (\ref{einc})
\be
\frac{t}{e^t-1}=1-\frac{1}{2}t+\sum_{k=1}^{\infty}(-1)^{k-1}B_k\frac{t^{2k}}{(2k)!}\; ,
\label{ber}
\ee

\noindent
obtaining ($B_2=\frac{1}{30}$):
\be
{\cal E}_{\rm vac}^{\rm inner}(\Lambda ,d)= -\frac{1}{4\pi\Lambda^3}-\frac{1}{2}\;\frac{\pi^2}{720d^3}+{\cal O}(\Lambda )\; ,
\label{einf}
\ee

\noindent
and thus
\be
{\cal E}_{\rm Casimir}^{\rm inner}=-\frac{1}{2}\;\frac{\pi^2}{720d^3}\; .
\ee

Two remarks are in order. The surface term $-\frac{1}{4\pi\Lambda^3}$ in (\ref{einf}) are absent for periodic b.c., because the latter allow the term $n=0$ in (\ref{ein}) which exactly cancels it. This explains the result of \cite{SWr92}. The external CE is {\it zero} due to (\ref{eigou}), (\ref{omou}) because, for the outer problem, $d$ in (\ref{einf}) is replaced by $R$, and thus in the limit $R\rightarrow \infty$ 
\be
{\cal E}_{\rm Casimir}^{\rm outer}=0\; .
\ee

\noindent
Finally,
\be
{\cal E}_{\rm Casimir}=-\frac{1}{2}\;\frac{\pi^2}{720d^3}\; .
\label{ecf}
\ee

\noindent
The above energy is one half of the result for the electromagnetic field, due to the summation over the two polarization states in the latter. Notice also that, in natural units, $\cal E$ is of order $(length)^{-3}$.

An amusing aspect of the present derivation is that it seems to depend on the choice (\ref{ck}), i.e., of an exponential cutoff in (\ref{ein}) and (\ref{einc}), which, due to (\ref{ber}), leads to (\ref{einf}). Consider now a general cutoff function (\ref{esp}). Omitting the volume term in (\ref{ein}), we may write
\be
{\cal E}_{\rm vac}^{\rm inner}(\Lambda ,d)=\lim_{n\rightarrow \infty}\frac{1}{8\pi}\sum_{m=1}^n g(m)\; ,
\label{eg}
\ee

\noindent
where
\ba
g(m)&=& \int_0^{\infty}du\; \sqrt{u+\left(\frac{m\pi}{d}\right)^2} \;C\left(\Lambda\sqrt{u+\left(\frac{m\pi}{d}\right)^2}\right)\nonumber \\ \nonumber\\
&=& \int_{\left(\frac{m\pi}{d}\right)^2}^{\infty}du\;\sqrt{u}\; C(\Lambda\sqrt{u})\; .
\eea

It is of interest to compute
{
\setcounter{let}{1}
\renewcommand{\theequation}{\arabic{equation}\alph{let}}
\ba
\frac{d}{\pi}g^{(1)}(m)&=&-2\left(\frac{m\pi}{d}\right)^2C\left(\Lambda\frac{m\pi}{d}\right)\; ; \label{g1}\\ \nonumber \\ \addtocounter{let}{1} \addtocounter{equation}{-1}
\frac{d}{\pi}g^{(2)}(m)&=&-4\frac{\pi}{d}\left(\frac{m\pi}{d}\right)C\left(\Lambda\frac{m\pi}{d}\right)-2\left(\frac{m\pi}{d}\right)^2\left(\frac{\Lambda\pi}{d}\right)C^{(1)}\left(\Lambda\frac{m\pi}{d}\right)\; ;\label{g2} \\ \nonumber \\ \addtocounter{let}{1} \addtocounter{equation}{-1}
\frac{d}{\pi}g^{(3)}(m)&=&-4\left(\frac{\pi}{d}\right)^2C\left(\Lambda\frac{m\pi}{d}\right)-8\frac{\pi}{d}\left(\frac{m\pi}{d}\right)\left(\frac{\Lambda\pi}{d}\right)C^{(1)}\left(\Lambda\frac{m\pi}{d}\right)\nonumber \\ \nonumber \\
&-&2 \left(\frac{m\pi}{d}\right)^2 \left(\frac{\Lambda\pi}{d}\right)^2C^{(2)}\left(\Lambda\frac{m\pi}{d}\right)\; ; \label{g3}  \; .
\eea
}

By \cite{Har49} (p. 326), under the following conditions (\ref{csum}) and (\ref{ckzero}) on $C$:
\be
\sum_{m=1}^ng(m)-\frac{2d}{\pi}\int^{\infty}_0dq\;q^3\;C(\Lambda q)+\frac{1}{2}g(0) \;\; \begin{array}{c} \\ _{n\rightarrow \infty}\end{array}\!\!\!\!\!\!\!\!\!\!\!\!\!\!\! \longrightarrow \; \Sigma_k\; ,
\label{gm}
\ee

\noindent
where
\be
\Sigma_k=-S_k(0)-\frac{1}{(2k+2)!}\int_0^{\infty}\psi_{2k+2}(t)g^{(2k+2)}(t)dt\; ,
\label{sig}
\ee

\noindent
and
\ba
\psi_k(x) =\phi_k(x) \; {\rm mod}\;1 \;\; \;&&({\rm i.e.,\; equal\; to} \;\phi_k(x) \;{\rm for}\; \\ 
&&0\leq x<1 \;{\rm with\; period}\; 1)\; ,\nonumber
\eea

\noindent
and $\phi_k$ are defined by
\be
t\frac{e^{xt}-1}{e^t-1}=\sum_{n=1}^{\infty}\phi_k(x)\frac{t^n}{n!}\; ,
\ee

\noindent
and
\be
S_k(0)=\sum_{r=1}^k (-1)^{r-1}\frac{B_r}{(2r)!}g^{(2r-1)}(0)\; .
\label{sk0}
\ee

\noindent
We changed the notation of \cite{Har49}: the $C_k$ on pg. 326 corresponds to our $\Sigma_k$. Notice that the second term in (\ref{gm}) corresponds to the subtraction of the vacuum term, which appears in a natural way as a necessary requirement in a purely mathematical context! The term $\frac{1}{2}g(0)$ contributes only to the $\Lambda$-dependent terms in the asymptotic series.

{\bf Theorem}$\;$ {\it Let the special cutoff function of type (\ref{esp}) satisfy, besides (\ref{noresp}), the conditions: $C$ is infinitely differentiable and its derivatives $C^{(k)}$ ($C^{(0)}\equiv C$) satisfy
\ba
&&\int^{\infty}C^{(k)}(x)dx <\infty \; ; \label{csum}\\ \nonumber \\
&&C^{(k)}(x)\begin{array}{c} \\ _{x\rightarrow \infty}\end{array}\!\!\!\!\!\!\!\!\!\!\!\!\!\!\! \longrightarrow \; 0 \; .\label{ckzero}
\eea

\noindent
Then, for Dirichlet (or Neumann) b.c. the $\Lambda$-independent term in (\ref{easym}) is the cutoff-independent part of the Ramanujan sum of the divergent series (\ref{ereg}) with $C_{\alpha}(\Lambda )\equiv 1$, and is therefore RI, i.e., independent of $C$.}

\vspace{0.2cm}

{\bf Remark}$\;$ $\Sigma_k$ ($k\geq 1$) is referred to as the $(\Re ,0)$ sum of the (divergent) series $\sum_{m=1}^{\infty}g(m)$, where $\Re$ refers to Ramanujan and $0$ to the reference point (the origin in our case). Usually (see, e.g., \cite{IZu80}, p. 138), the result is presented informally without the important last term in (\ref{sig}), and assuming that $C$ satisfies $C^{(k)}(0)=0$ for all $k\geq 1$, besides (\ref{noresp}), which is not satisfied by the special choice (\ref{ck}) (see, however, \cite{PMG86} for a nice approach to the subject). 

{\it Proof}. The fact that $\Sigma_k$ is independent of $k$ for $k\geq 1$ follows from \cite{Har49} (pp 326 ff). Choose $k=2$. By (\ref{sig})-(\ref{sk0}),
\be
\Sigma_2=-\frac{B_1}{2}g^{(1)}(0)+\frac{B_2}{24}g^{(3)}(0)-\frac{1}{6!}\int_0^{\infty}\psi_6(t)g^{(6)}(t)dt\; .
\label{sig2}
\ee

\noindent
Putting (\ref{g1}), (\ref{g3}) and (\ref{noresp}) into (\ref{sig2}), we find
\be
\Sigma_2=-\frac{B_2}{6}\left(\frac{\pi}{d}\right)^3 +{\cal O}(\Lambda^2)\; ,
\label{rsig2}
\ee

\noindent
which leads to (\ref{ecf}) by (\ref{eg}). The term ${\cal O}(\Lambda^2)$ in (\ref{rsig2}) comes from $g^{(6)}$, making the change of variable $t'=\frac{\Lambda\pi}{d}t$ in the integral in (\ref{sig2}) and taking into account that $\psi_k$ is ${\cal O}(1)$.

What if we choose $k=1$? By (\ref{sk0}) and (\ref{g1}), $S_1(0)=0$, but, in (\ref{sig}), we still have the second term
\be
\Sigma_1=-\frac{1}{24}\int_0^{\infty}\psi_4(t)g^{(4)}(t)dt\; .
\label{sig1}
\ee

\noindent
We use the recurrence (\cite{Har49}, 13.2.13)
\be
\psi^{(1)}_{2m+1}=(2m+1)\left\{\psi_{2m}+(-1)^{m-1}B_m\right\}\; ,
\ee

\noindent
with $m=2$, obtaining
\be
\psi_4-B_2=\frac{1}{5}\psi_5^{(1)}\; ,
\ee

\noindent
which we insert in (\ref{sig1}), getting
\be
\Sigma_1=-\frac{1}{24}\int_0^{\infty}\frac{\psi^{(1)}_5(t)}{5}g^{(4)}(t)dt-\frac{1}{24}B_2\int_0^{\infty}g^{(4)}(t)dt \; .
\label{sig1c}
\ee

\noindent
Integration by parts in the first term on the r.h.s. of (\ref{sig1c}) and use of (\ref{g3}) in the second term yield (using $\psi_n(0)=0$)
\be
\Sigma_1=\frac{1}{120}\int_0^{\infty}\psi_5(t)g^{(5)}(t)dt+\frac{B_2}{24}g^{(3)}(0) \; .
\label{sig1c1}
\ee

\noindent
A further integration by parts using the recurrence (\cite{Har49}, 13.2.13)
\be
\psi_{2m}^{(1)}= 2m \psi_{2m-1}\; ,
\ee

\noindent
brings (\ref{sig1c1}) to the form (\ref{sig2}). We have thus proved
\be
\Sigma_k=-\frac{B_2}{6}\left(\frac{\pi}{d}\right)^3 +{\cal O}(\Lambda^2)\; ,
\label{rsigk}
\ee

\noindent
for all $k\geq 1$ (the present argument is easily generalized). Thus, for parallel plates and Dirichlet b.c. the $\Lambda$-independent term in the asymptotic series (\ref{easym}) is regularization independent and is the $(\Re ,0)$ sum of the divergent series (\ref{eg}). Neumann b.c. yield the same result.

\section{\large The Sphere}
\label{sphere}

The Casimir effect for b.c. on a sphere was first considered in the classic paper by Boyer \cite{Boy68} and since it has been considered from diverse viewpoints: source theory \cite{MRS78}, multiple scattering \cite{BDu78}, dimensional dependence of the effect \cite{BMi94} as well as an improved mode summation method \cite{NPi98, BHa98} (see also \cite{BEK96}). In \cite{Kli00} it is shown how a natural subtraction method ensures convergence of the mode sum and in \cite{BKV99} RI for the ball has been proved (for a more detailed reference list see \cite{Lam99} and \cite{PMG86}).

Here we will to reconsider the CE for a massless scalar field subjected to Dirichlet b.c. on a sphere in the light of the above developed theory. We will consider the original sphere, of radius $a$, embedded in a concentric greater sphere of radius $R>a$. As it is well-known, for the sphere it is convenient to consider the inner and outer regions together in order to avoid logarithmic contributions for the CE. So, taking into account the $(2l+1)$-fold degeneracy of each eigenvalue, we have 
\begin{equation}
E_{\rm vac}= \sum_{l=0}^{\infty}(l+\frac{1}{2})\sum_{n=1}^{\infty}\omega_{nl}\;C_{nl}(\Lambda )\; ,
\label{eregs}
\end{equation}

\noindent
where $\omega_{nl}$ are the eigenfrequencies. The sum over $n$ in (\ref{eregs}) can be changed into an integral by using the Cauchy theorem \cite{NPi98, BHa98, BEK96}. Here we will follow \cite{BHa98} with an crucial difference: the cutoff functions used in \cite{BHa98}, while appropriate to treat the eletromagnetic field, do not render the integrals well-defined in the massless scalar field case, so that we will adopt\footnote{More general cutoffs of the above type have been used by C. R. Hagen in a different context \cite{Hag01}.}
\be 
C_{nl}(\Lambda)=e^{-\Lambda(\nu /a+\omega_{nl})}\; ,
\label{cmod}
\ee

\noindent
for the cutoff functions, which satisfies the normalization condition (\ref{normaliz}). Besides this, it is important to analyse the asymptotic behaviour of $E_{\rm vac}$ based on more general cutoff functions.

Then, we can rewrite (\ref{eregs}) as \cite{NPi98, BHa98}
\begin{equation}
E_{\rm vac}= -\frac{1}{a}\sum_{l=0}^{\infty} Q_l\; ,
\label{eql}
\end{equation}

\noindent
with ($\nu =l+\frac{1}{2}$)
\be
Q_l = \frac{\nu^2}{\pi} e^{-\Lambda\nu /a}{\rm Re}\; e^{-i \varphi}\int_0^{\infty}y\exp \{-i\nu\frac{\Lambda}{a}ye^{-i\varphi}\} \frac{d}{dy}\ln f_l(i\nu ye^{-i\varphi})\; dy \; ,
\label{ql}
\ee

\noindent
where $\varphi$ is an (small) angle which orientates the contour of integration with respect to the imaginary axis of $z$ (see \cite{BHa98}), and
\be
f_l(i z) = -\frac{1}{ z}I_{\nu}( z)K_{\nu}( z)\; .
\label{fli}
\ee

\noindent
Now using the uniform asymptotic expansions for the Bessel functions $I_{\nu}$ and $K_{\nu}$ \cite{ASt65} we can obtain an asymptotic expansion for $Q_l$ which is valid for large orders. Then, in general, we can rewritte $E_{\rm vac}$ as \cite{NPi98}
\begin{equation}
E_{\rm vac} = E_{\rm asym}-\frac{1}{a} \sum_{l=0}^{n}\Delta Q_l \; ,
\label{edq}
\end{equation}

\noindent
where $E_{\rm asym}$ stands for the expression obtained from (\ref{eql})-(\ref{ql}) by using the asymptotic expansions for the Bessel functions \cite{ASt65}, $\Delta Q_l=Q_l-Q_l^{\rm asym}$ and $n$ is such that for $l>n$ the asymptotic expansion $ Q_l^{\rm asym}$ affords a good approximation for $ Q_l $ (i.e., $\Delta Q_l \simeq 0$ for $l>n$).

Then, we obtain (after performing a rotation of the integration contour $ye^{-i\varphi}\rightarrow y$)
\ba
E_{\rm asym}\!\!\! &=& - \frac{2}{\pi}\frac{a^2}{\Lambda^3} {\rm Re} \int_0^{\infty} \frac{y}{(1+iy)^3}\frac{d}{dy} \ln t \; dy \nonumber \\ \nonumber  \\
&-& \frac{1}{\pi } \frac{1}{\Lambda} {\rm Re}\int_0^{\infty} \frac{y}{(1+iy)} \frac{d}{dy}\alpha (t) \; dy  \label{e} \\ \nonumber  \\
&-&\!\! \frac{1}{\pi a}\zeta (2, \frac{1}{2})\; {\rm Re} \int_0^{\infty} y \frac{d}{dy}\left[\beta (t)-\frac{1}{2}\alpha^2(t)\right] dy + {\cal O}(\Lambda ) \; ,\nonumber
\eea

\noindent
where $\zeta (s, a)=\sum_{l=0}^{\infty}(l+a)^{-s}$ is the Hurwitz zeta function. From this expression must be clear why we have introduced the cutoff functions (\ref{cmod}) rather than $e^{-\Lambda\omega_{nl}}$ used in \cite{BHa98}. Namely, in the absence of the term $e^{-\Lambda\nu /a}$ in (\ref{cmod}) the first integral in (\ref{e}) would have a non-integrable singularity in the origin, but all integrals are well-defined if we adopt (\ref{cmod}). 

From (\ref{e}) we have ($\zeta (2, \frac{1}{2}) = \frac{\pi^2}{2}$)
\ba
E_{\rm asym} = -\frac{a^2}{8\;\Lambda^3} - \frac{5}{1024\;\Lambda} + \frac{35\pi^2}{65536\; a} + {\cal O}(\Lambda ) \; .
\label{eass}
\eea

Now, it remains to calculate $\sum_{l=0}^n\Delta Q_l$ in (\ref{edq}). Notice that in this term the sum is finite and we do not have any divergence. Then, since $\varphi >0$ may be considered a small angle ($\sin \varphi >0$ and $\cos \varphi >0$) we may integrate (\ref{ql}) by parts and perform a rotation of the integration contour ($ye^{-i\varphi}\rightarrow y$) to obtain
\be
Q_l= -\frac{\nu}{\pi}\int_0^{\infty}dy \ln \left[ 2yI_{\nu}(y)K_{\nu}(y)\right] + {\cal O}(\Lambda)\; ,
\label{qnes}
\ee

\noindent
which is nothing but the $Q_l$ in ref. \cite{NPi98} (except for a sign). So we may take advantage of the numerical results in \cite{NPi98} for this expression.

Analogously, we may obtain a expression for $Q_l^{\rm asym}$, appropriate for when there is no infinite summation, given by 
\ba
Q_l^{\rm asym}\!\!\! &=&\!\!\!-\frac{\nu^2}{\pi } \int_0^{\infty} dy \frac{d}{dy}\ln \left[ \frac{y}{\sqrt{1+y^2}}\right]- \frac{1}{\pi}\int_0^{\infty} dy \;\alpha (t) \nonumber \\ \nonumber  \\
&-& \frac{1}{\pi \nu^2}\int_0^{\infty} dy \left[\beta (t)-\frac{1}{2}\alpha^2(t)\right] + {\cal O}(\Lambda)\; ,
\eea

\noindent
which after integration yields
\be
Q_l^{\rm asym}=\frac{\nu^2}{2} + \frac{1}{128} - \frac{35}{32768\; \nu^2}+\cdots\; .
\label{qnu}
\ee

\noindent
Then, we can take $n=4$ in (\ref{edq}) as a good approximation (see \cite{NPi98}) obtaining
\be
E_{\rm vac} = -\frac{a^2}{8\Lambda^3}-\frac{5}{1024\Lambda}+ \frac{0.002819}{a}+\cdots\; ,
\label{easysp}
\ee

\noindent
which yields $a^{\rm sphere}_4\simeq 0.002819$ for the coefficient of the $\Lambda$-independent term in the asymptotic series (\ref{easym}) for $E$. Therefore the CE is 
\be
E_{\rm Casimir}= \frac{a_4^{\rm sphere}}{a}\simeq \frac{0.002819}{a}\; ,
\label{cashell}
\ee

\noindent
and by (\ref{psin}) we see that the Casimir force for massless scalar field with Dirichlet b.c. on a sphere is repulsive. This result was obtained with greater precision in \cite{BMi94} (also see \cite{BHa98, NPi98, LRo96}).

While the numerical result provided by (\ref{cashell}) is not new, the above calculation illustrates the fact that when we use more general cutoffs like (\ref{cmod}) (which in the present case is mandatory) we are faced with an asymptotic series in $\Lambda$ for $E_{\rm vac}$, see (\ref{easym}) and (\ref{easysp}). Then, the method discussed above provides an unambiguous way to identify the CE.

\section{\large The Interior Problem for the Cube}
\label{cinner}

Consider now a cube $K$ of side $L$, with Dirichlet b.c. (Neumann b.c. may be handled analogously). The normalized eigenfunctions and eigenvalues of $(-\Delta)^{\frac{1}{2}}$ are 
\ba
&&u_{n_1n_2n_3}(\vec{x})=\left(\frac{2}{L}\right)^{\frac{3}{2}}\sin \left(\frac{n_1\pi x_1}{L}\right)\sin \left(\frac{n_2\pi x_2}{L}\right)\sin \left(\frac{n_3\pi x_3}{L}\right)\; ; \nonumber \\ \label{cube} \\
&&(-\Delta)^{\frac{1}{2}}u_{n_1n_2n_3}(\vec{x})=\frac{\pi}{L}|\vec{n}|u_{n_1n_2n_3}(\vec{x})\; ; \hspace{0.5cm} |\vec{n}|=(n_1^2+n_2^2+n_3^2)^{\frac{1}{2}} \; .\nonumber 
\eea

\noindent
where $n_i=1,2,\cdots$ ($i=1,2,3$).

We consider 
\be
E_{\rm vac}(\Lambda )=\int_Kd^3 x\; H(\vec{x},\Lambda )\; .
\ee

\noindent
By (\ref{hmom}),
\be
E_{\rm vac}(\Lambda )=\frac{1}{2}\frac{\partial}{\partial\Lambda}\left\{\frac{L^3}{(2\pi )^3}\int d^3k\;e^{-\Lambda|\vec{k}|}-\sum_{\vec{n}}e^{-\Lambda\omega_{\vec{n}}}\right\}\; .
\ee

\noindent
By (\ref{cube}), $\omega_{\vec{n}}=\frac{\pi}{L}|\vec{n}|$ and hence
\be
E_{\rm vac}(\Lambda )=-\frac{3}{2\pi^2}L^3\Lambda^{-4} -\frac{1}{16}\frac{\partial}{\partial\Lambda}\left[\sum_{\vec{n}\in \mathbb{Z}^3}e^{-a|\vec{n}|}-3\sum_{\vec{n}\in \mathbb{Z}^2}e^{-a|\vec{n}|}+3\sum_{n\in \mathbb{Z}}e^{-a|n|}-1 \right]\; ,
\label{esums}
\ee

\noindent
where
\be
a=\frac{\pi}{L}\Lambda\; .
\ee

The last sums in (\ref{esums}) are due the fact that, because of (\ref{cube}), the planes $n_1=0$, $n_2=0$, and $n_3=0$ have to be excluded from the sum over $\mathbb{Z}^3$ because they lead to eigenfunctions which are zero. For the same reason the axes $n_1=n_2=0$, $n_1=n_3=0$, $n_2=n_3=0$ and the origin be excluded. Exclusion of the three planes (the term $-3\sum_{\vec{n}\in \mathbb{Z}^2}e^{-a|\vec{n}|}$ in (\ref{esums})) corresponds to excluded each axis twice instead of only once. The third term compensates for this, while the last one excludes the origin.

A method of calculation of the lattice sums in (\ref{esums}) is through the Poisson summation formula
\be
\sum_{\vec{n}\in \mathbb{Z}^3}f(\vec{n})=\sum_{\vec{m}\in \mathbb{Z}^3}C_{\vec{m}}\; ,
\label{pois}
\ee

\noindent
where $C_{\vec{m}}$ are the Fourier coefficients of $f$:
\be
C_{\vec{m}}=\int d^3x\; e^{-2\pi i\vec{m}\cdot\vec{x}}f(\vec{x})\; .
\ee

\noindent
See also ref. \cite{RZV77}. Applying (\ref{pois}) to (\ref{esums}), we find
\ba
E_{\rm vac}(\Lambda )\!\!&=&-\frac{3}{2\pi^2}L^3\Lambda^{-4} +\frac{3}{2\pi^2}L^3\Lambda^{-4} -\frac{3}{4\pi}L^2\Lambda^{-3}+\frac{3}{8\pi}L\Lambda^{-2} \nonumber \\ \nonumber \\
&-&\!\!\!\frac{\pi^2}{2L}\!\!\sum_{  \stackrel{\scriptscriptstyle \vec{m}\in \mathbb{Z}^3 }{\scriptscriptstyle \vec{m}\neq \vec{0}}}\left[\!\left(\frac{\pi\Lambda}{L}\right)^2\!\!+4\pi^2|\vec{m}|^2\right]^{-2} \!\!\!\!\!+ \frac{2\pi^4}{L}\!\left(\frac{\Lambda}{L}\right)^2\!\!\sum_{  \stackrel{\scriptscriptstyle \vec{m}\in \mathbb{Z}^3 }{\scriptscriptstyle \vec{m}\neq \vec{0}}}\left[\!\left(\frac{\pi\Lambda}{L}\right)^2\!\!\!+4\pi^2|\vec{m}|^2\right]^{-3} 
\nonumber \\ \label{esoma}  \\
&+&\!\!\!\!\frac{3\pi^2}{8L}\!\!\sum_{  \stackrel{\scriptscriptstyle \vec{m}\in \mathbb{Z}^2 }{\scriptscriptstyle \vec{m}\neq \vec{0}}}\!\left[\!\left(\frac{\pi\Lambda}{L}\right)^2\!\!\!+4\pi^2|\vec{m}|^2\right]^{-\frac{3}{2}} \!\!\!\!\!-\frac{9\pi^4}{8L}\!\left(\frac{\Lambda}{L}\right)^2\!\!\sum_{  \stackrel{\scriptscriptstyle \vec{m}\in \mathbb{Z}^2 }{\scriptscriptstyle \vec{m}\neq \vec{0}}}\!\left[\!\left(\frac{\pi\Lambda}{L}\right)^2\!\!\!+4\pi^2|\vec{m}|^2\right]^{-\frac{5}{2}} \nonumber \\ \nonumber \\
&-&\!\!\!\frac{3\pi}{8L}\sum_{  \stackrel{\scriptscriptstyle m\in \mathbb{Z} }{\scriptscriptstyle m\neq 0}}\left[\left(\frac{\pi\Lambda}{L}\right)^2\!\!+4\pi^2m^2\right]^{-1}\!\!\!\!+ \frac{3\pi^3}{4L}\left(\frac{\Lambda}{L}\right)^2\sum_{  \stackrel{\scriptscriptstyle m\in \mathbb{Z} }{\scriptscriptstyle m\neq 0}}\left[\left(\frac{\pi\Lambda}{L}\right)^2\!\!+4\pi^2m^2\right]^{-2}\!\! . \nonumber
\eea

\noindent
We now expand the sums $\sum_{  \vec{m}\neq \vec{0} }$ in (\ref{esoma}) in the following way:
\be
\left[\left(\frac{\pi\Lambda}{L}\right)^2\!\!+4\pi^2|\vec{m}|^2\right]^{-s}= \left(4\pi^2|\vec{m}|^2\right)^{-s}\left(1-\frac{s\Lambda^2}{4L^2|\vec{m}|^2}+\cdots \right)\; .
\label{exp}
\ee

\noindent
The unit term in (\ref{exp}) yields a contribution of type $a_4L^{-1}$ in (\ref{easym}), the remaining terms provide the rest of the asymptotic series in (\ref{easym}) consisting of positive powers of $\Lambda$. We thus find
\be
a_4=-\frac{1}{32\pi^2}\sum_{  \stackrel{\scriptscriptstyle \vec{m}\in \mathbb{Z}^3 }{\scriptscriptstyle \vec{m}\neq \vec{0}}}|\vec{m}|^{-4}+\frac{3}{64\pi}\sum_{  \stackrel{\scriptscriptstyle \vec{m}\in \mathbb{Z}^2 }{\scriptscriptstyle \vec{m}\neq \vec{0}}}|\vec{m}|^{-3}-\frac{3}{32\pi}\sum_{  \stackrel{\scriptscriptstyle m\in \mathbb{Z} }{\scriptscriptstyle m\neq 0}}m^{-2}\; .
\label{a4cube}
\ee

\noindent
The last sum above is nothing but $2\zeta (2)$, where $\zeta$ stands for the Riemann zeta function, and the second one may be rewritten as the product of two independent sums by means of $\sum_{  \stackrel{\scriptscriptstyle \vec{m}\in \mathbb{Z}^2 }{\scriptscriptstyle \vec{m}\neq \vec{0}}}|\vec{m}|^{-s}= 4\zeta \left(\frac{s}{2}\right) \beta \left(\frac{s}{2}\right)$ (see, e.g., \cite{AWo82, LCL97}), where $\beta (s) = \sum_{j=0}^{\infty}\frac{(-1)^j}{(2j+1)^s}$. Then, using the result of Lukosz \cite{Luk71} for the first sum in (\ref{a4cube}) we obtain
\be
a_4=-0.0157322\ldots \; , 
\label{a4n}
\ee

\noindent
which is in accordance with the result obtained numerically in ref. \cite{AWo82} (in fact we have obtained $a_4$ to a higher accuracy than shown). In addition, from (\ref{pin}), the inner pressure is
\be
p_C^{\rm inner}(L)=\frac{a_4}{3L^4}\; .
\label{pcin}
\ee

By (\ref{a4n}) and (\ref{pcin}) we see that the force due to the interior is attractive. The repulsive character of the sphere (section \ref{sphere}) suggests, however, that the same is true for the cube. This fact alone shows that this sign, if true, {\it must} be entirely due to the exterior force, a subtle problem to which we now turn.

\section{\large The External Problem for the Cube}
\label{couter}

As remarked above, it is of great interest to consider also the outer problem for the cube. We will consider the cube $K_L$ of side $L$ concentric with a cube $K_M$, of side $M$, from which $K_L$ is a subset ($M>L$ and $M$ eventually goes to infinity at the end of calculation) and impose Dirichlet b.c. on $K_L$ as well as $K_M$ (see section 1). Unfortunately, the solution of the external Casimir problem for the cube with Dirichlet b.c. cannot be constructed out of the functions of the form (\ref{cube}), because the continuity conditions on several planes cannot be satisfied simultaneously. However, the form of solutions (\ref{cube}), which are naturally adapted to the internal geometry of the cube, suggest splitting the region $K_M\backslash K_L$ into 26 subregions bounded by the planes containing the faces of the cube. We may require the $u_n(\vec{x})$ to vanish on the boundaries of these subregions, including the original requirement of vanishing on the faces of the internal and external cubes. If we do so, the resulting problem is explicitly solvable in terms of the set (\ref{cube}). Of course, this Ansatz introduces additional stresses in the region $K_M\backslash K_L$. We shall comment on these restrictions at the end of this section.

Then we have that the 26 subregions which compose $K_M\backslash K_L$ are of three topologically distinct kinds (with both cubes centered in the origin): 1) a rectangular box with two sides $L$ and one $\frac{M-L}{2}$ -- with multiplicity 6; 2) a rectangular box with two sides $\frac{M-L}{2}$ and one $L$ (with multiplicity 12) -- the contribution of the edges; 3) a cube of sides $\frac{M-L}{2}$ (with multiplicity 8) -- the contribution of the corners. The Casimir energy of each of these regions can be obtained along the same lines of the calculation above outlined for the inner cube (see \cite{RZV77}). Then we obtain that the regions of type 3) do not contribute, i.e., there are no contributions of the corners, either to the energy or the pressure, in the limit $M\rightarrow \infty$. The total contribution of the regions of type 1) is
\ba
E_1(L, M)&=&-\frac{3L^2(M-L)}{32\pi^2}\sum_{ \stackrel{\scriptscriptstyle \vec{m}\in \mathbb{Z}^3 }{\scriptscriptstyle \vec{m}\neq \vec{0}}}\frac{1}{\left[ m_1^2L^2+m_2^2L^2+m_3^2\frac{(M-L)^2}{4}\right]^2} \nonumber \\ \nonumber \\
&+& \frac{3L(M-L)}{32\pi} \sum_{  \stackrel{\scriptscriptstyle \vec{m}\in \mathbb{Z}^2 }{\scriptscriptstyle \vec{m}\neq \vec{0}}}\frac{1}{\left[m_1^2L^2+m_2^2\frac{(M-L)^2}{4}\right]^{\frac{3}{2}}} \\ \nonumber \\
&+& \frac{3}{32\pi L}\sum_{  \stackrel{\scriptscriptstyle \vec{m}\in \mathbb{Z}^2 }{\scriptscriptstyle \vec{m}\neq \vec{0}}}\frac{1}{\left[m_1^2+m_2^2\right]^{\frac{3}{2}}} - \frac{\pi}{16}\left( \frac{2}{L}+\frac{2}{M-L}\right) \; ,\nonumber
\eea

\noindent
and for the regions of type 2) the total contribution is
\ba
E_2(L, M)&=&-\frac{3L(M-L)^2}{32\pi^2}\sum_{  \stackrel{\scriptscriptstyle \vec{m}\in \mathbb{Z}^3 }{\scriptscriptstyle \vec{m}\neq \vec{0}}}\frac{1}{\left[ m_1^2L^2+m_2^2\frac{(M-L)^2}{4}+m_3^2\frac{(M-L)^2}{4}\right]^2} \nonumber \\ \nonumber \\
&+& \frac{3L(M-L)}{16\pi} \sum_{  \stackrel{\scriptscriptstyle \vec{m}\in \mathbb{Z}^2 }{\scriptscriptstyle \vec{m}\neq \vec{0}}}\frac{1}{\left[m_1^2L^2+m_2^2\frac{(M-L)^2}{4}\right]^{\frac{3}{2}}} \label{edges} \\ \nonumber \\
&+& \frac{3}{8\pi (M-L)}\sum_{  \stackrel{\scriptscriptstyle \vec{m}\in \mathbb{Z}^2 }{\scriptscriptstyle \vec{m}\neq \vec{0}}}\frac{1}{\left[m_1^2+m_2^2\right]^{\frac{3}{2}}} - \frac{\pi}{8}\left( \frac{1}{L}+\frac{4}{M-L}\right) \; .\nonumber
\eea

\noindent
Then, by (\ref{pou}) and taking into account that $dV_{\rm outer}= 6dV_1 +12dV_2 +8dV_3$, where $V_i$ is the volume of type $i$) region, and that in the thermodynamic limit $dV_3\rightarrow 0$ ($dV_3/V_3 \propto 1/M$), we obtain
\be
p_C^{\rm outer}(L)=-\lim_{M\rightarrow \infty}\frac{1}{3M(M-2L)}\frac{\partial}{\partial L}\left[ E_1(L,M)+E_2(L,M)\right]\; .
\label{limpou}
\ee

\noindent
It may also be verified explicitly that regions of type 3) do not contribute to the energy, or to the pressure, in the thermodynamic limit. From (\ref{limpou}) we see that type 1) regions do not contribute to the outer pressure, while the edges contribution is given only by the first term in (\ref{edges}): 
\be
p_C^{\rm outer}(L)=  \lim_{M\rightarrow \infty}\left\{ \frac{\partial}{\partial L}\left( \frac{L}{32\pi^2}\sum_{  \stackrel{\scriptscriptstyle \vec{m}\in \mathbb{Z}^3 }{\scriptscriptstyle m_3\neq 0}}\frac{1}{\left[ m_1^2L^2+m_2^2\frac{(M-L)^2}{4}+m_3^2\frac{(M-L)^2}{4}\right]^2}  \right) \right\} \; .
\label{pcou} 
\ee

\noindent
It follows from (\ref{pcou}) that (see later):
\be
p_C^{\rm outer}(L)=\frac{1}{32\pi^2}\left( \sum_{  \stackrel{\scriptscriptstyle m\in \mathbb{Z} }{\scriptscriptstyle m\neq 0}}m^{-4}\right) \frac{\partial}{\partial L}L^{-3} = -\frac{\pi^2}{480}L^{-4}\; . 
\label{respou} 
\ee

There is no contribution in the thermodynamic limit to (\ref{pcou}) from the sum over $m_2 \neq 0$ or $m_3\neq 0$, or both. Indeed, the term $m_1 =0$ in (\ref{pcou}) does not contribute as $M\rightarrow \infty$, as one sees easily, and
\be
\sum_{ \stackrel{ \stackrel{\scriptscriptstyle \vec{m}\in \mathbb{Z}^3 }{\scriptscriptstyle m_1\neq 0 }}{\scriptscriptstyle m_2\neq 0, m_3\neq 0} } \!\!\!\!\!\!\! \frac{1}{\left[\! m_1^2L^2\! +\! m_2^2\frac{(M\! -\! L)^2}{4}\! +\! m_3^2\frac{(M\! -\! L)^2}{4}\!\right]^2}  \leq  \left(\!\frac{2}{L^7(M\! -\! L)}\!\right)^{\frac{1}{2}}\!\!\!\!\! \sum_{ \stackrel{ \stackrel{\scriptscriptstyle \vec{m}\in \mathbb{Z}^3 }{\scriptscriptstyle m_1\neq 0 }}{\scriptscriptstyle m_2\neq 0, m_3\neq 0} }\!\!\!\!\!\!\! \frac{1}{\left[ m_1^2\! +\! m_2^2\! +\! m_3^2 \right]^{\frac{7}{4}}}\; . 
\label{est1} 
\ee

\noindent
If only $m_2\neq 0$ (or $m_3\neq 0$)
\be
\sum_{  \stackrel{\scriptscriptstyle \vec{m}\in \mathbb{Z}^2 }{\scriptscriptstyle m_1\neq 0, m_2\neq 0}}\frac{1}{\left[ m_1^2L^2+m_2^2\frac{(M\! -\! L)^2}{4} \right]^2}  \leq  \left(\frac{2}{L^3(M-L)}\right)\sum_{  \stackrel{\scriptscriptstyle \vec{m}\in \mathbb{Z}^2 }{\scriptscriptstyle m_1\neq 0, m_2\neq 0}}\frac{1}{\left[ m_1^2+m_2^2\right]^{\frac{3}{2}}} \; , \label{est2}
\ee

\noindent
for $L\leq 1$, $M>L$. It may be checked that these bounds suffice to show that the contributions of the above sums to the limit in (\ref{pcou}) is zero. The only surviving term in the sum in (\ref{pcou}) is thus $m_2=m_3=0$, which leads to (\ref{respou}).

It is most important to note that the edges' contribution to the Casimir pressure is greater than the inner pressure in absolute value: by (\ref{pcin}), (\ref{respou}) and (\ref{p})
\be
p_C(L)=(-0.005244+\frac{\pi^2}{480})L^{-4} = 0.015317\; L^{-4}
\ee

\noindent
The Casimir pressure is thus repulsive, and the net result is that edge effects {\it determine the sign of the force}. We shall return to this point in the conclusion.

We now comment on our Ansatz. We have replaced the original Hilbert space of $L^2$-functions with Dirichlet b.c. in the inner and outer boundaries with a direct sum of 26 spaces, upon introduction of additional Dirichlet b.c. on planes which are extensions of the cube's faces to the region $K_M\backslash K_L$. In this region all extra stresses are {\it parallel} to the cube's faces, and for this reason the Casimir pressure is insensitive to their inclusion. We have verified this assertion in appendix A by introducing and extra Dirichlet plane orthogonal to a system of parallel plates in the inner region. The proof generalizes to an arbitrary finite number of such planes provided they are placed symmetrically to some plane orthogonal to the $z$-axis, thus not introducing an extraneous length in the original problem. Since the parallel plates are a soluble ``limiting case'' of the cube, we (strongly) believe that our Ansatz provides the exact solution for the cube.

\section{\large Conclusion and Open Problems}
\label{conclusion}

In this paper we have introduced a mathematically precise framework for the Casimir pressure, by associating it to the cutoff-independent part of the Ramanujan sum of the (divergent) series for the Casimir energy. Our ideas have precursors in \cite{Die88} and \cite{SWr92}. We illustrated the framework by parallel plates, the sphere and the interior problem of the cube.

In section \ref{couter} we introduced an Ansatz to calculate the exterior Casimir pressure for the cube. We discussed why we (strongly) believe that it is the exact solution for the cube. If our conjecture is right, the calculation of section \ref{couter} also provides an explanation for the sign of the force: it is due to a competition betweeen the inner and outer pressures, in which the latter is positive and larger than the former in absolute value, because, as remarked in section \ref{couter}, the thermodynamic limit selects a set of modes different from the inner ones\footnote{See also \cite{LRo96} for a discussion (different of ours) of the different roles of the inner and outer modes of the Casimir problem}, with a large positive contribution from the edges. The edges reflect the passage from the infinitely extended parallel plates to a compact region, i.e., by folding. If this folding were smooth, i.e., for any smooth approximation to the cube, it would be accompanied by nonzero {\it curvature}. At the other extreme -- {\it uniform} nonzero curvature -- we have the sphere. Here, however, curvature effects appear less directly, reflecting themselves in the appearance of the Neumann functions in the external problem. It is an interesting open problem to understand more clearly the role of curvature (of various kinds, e.g. Riemannian, the mean and Gaussian curvatures) in the Casimir effect for general compact manifolds with boundary (see also \cite{Die88}).

\section*{\large Acknowledgements}

One of us (W.W.) should like to thank G. Scharf for having introduced him to this subject, and posing some of the questions which are only partially answered in this paper. The authors thank Dr. K. Kirsten for information regarding ref. \cite{BEK96}. L. A. M. thanks FAPESP for full support (grant No. 99/04079-1). W. F. W. was supported in part by CNPq.



\section*{\large Appendix A}

\newcounter{apend}
\setcounter{apend}{1}
\renewcommand{\theequation}{\Alph{apend}.\arabic{equation}}
\setcounter{equation}{0}

In this appendix we consider the problem of the parallel plates with an orthogonal (Dirichlet) plane introduced in the inner region and prove that the Casimir {\it pressure} on the plates is the same that for the problem without the Dirichlet plane (see section \ref{plates}).

Consider the two parallel plates placed at $z=0$ and $z=d$, and the inner Dirichlet plane at $y=0$, $0\leq z\leq d$. The outer problem is the same which we worked out in section \ref{plates}, and it does not contribute. The inner Casimir problem is now split into two regions: $K_1^{\rm inner}=\{ \vec{x} \in \mathbb{R}\times [0,\infty )\times [0,d]\}$ and $K_2^{\rm inner}=\{ \vec{x} \in \mathbb{R}\times (-\infty ,0]\times [0,d]\}$. In order to calculate the Casimir energy of these regions it is necessary to consider a finite region in the plane $x$-$y$ with area $A=L_1L_2$ (the whole area of the plates), and take the limit $L_1,L_2\rightarrow \infty$ at the end. Then, both $K_1^{\rm inner}$ and $K_2^{\rm inner}$ are given by boxes of sides $L_1$ (along $x$-axis), $L_2/2$ (along $y$-axis) and $d$, so that we can proceed in the same way that in section \ref{cinner}, obtaining for the $\Lambda$-independent term of the series (\ref{easym}) for the region $K_1^{\rm inner}$:
\ba
E_1^{\rm inner}(L_1,L_2,d)&=&-\frac{L_1L_2d}{64\pi^2}\sum_{ \stackrel{\scriptscriptstyle \vec{m}\in \mathbb{Z}^3 }{\scriptscriptstyle \vec{m}\neq \vec{0}}}\frac{1}{\left[ m_1^2L_1^2+m_2^2\frac{L_2^2}{4}+m_3^2d^2\right]^2} \nonumber \\ \nonumber \\
&+& \frac{L_1L_2}{128\pi} \sum_{  \stackrel{\scriptscriptstyle \vec{m}\in \mathbb{Z}^2 }{\scriptscriptstyle \vec{m}\neq \vec{0}}}\frac{1}{\left[m_1^2L_1^2+m_2^2\frac{L_2^2}{4}\right]^{\frac{3}{2}}} + \frac{L_1d}{64\pi} \sum_{  \stackrel{\scriptscriptstyle \vec{m}\in \mathbb{Z}^2 }{\scriptscriptstyle \vec{m}\neq \vec{0}}}\frac{1}{\left[m_1^2L_1^2+m_2^2d^2\right]^{\frac{3}{2}}} \nonumber \\ \nonumber \\
&+& \frac{L_2d}{128\pi}\sum_{  \stackrel{\scriptscriptstyle \vec{m}\in \mathbb{Z}^2 }{\scriptscriptstyle \vec{m}\neq \vec{0}}}\frac{1}{\left[m_1^2\frac{L_2^2}{4}+m_2^2d^2\right]^{\frac{3}{2}}} - \frac{\pi}{96}\left( \frac{1}{L_1}+\frac{2}{L_2}+ \frac{1}{d}\right) \; , \label{e1in}
\eea

\noindent
and, obviously, $E_2^{\rm inner}(L_1,L_2,d)$ has the same form.

Now we can calculate the inner pressure by means of (\ref{pin}), where $V_{\rm inner}$ must be taken as the whole interior volume: $V_{\rm inner} =L_1L_2d$. Thus, we have
\ba
p_C(d)&=& p_C^{\rm inner}(d)= \lim_{L_1,L_2\rightarrow \infty}\left(-\frac{\partial E_1^{\rm inner}(L_1,L_2,d)}{\partial V_{\rm inner}}  -\frac{\partial E_2^{\rm inner}(L_1,L_2,d)}{\partial V_{\rm inner}} \right) \nonumber \\ \nonumber \\
&=& \lim_{L_1,L_2\rightarrow \infty}\left(-\frac{2}{L_1L_2}\frac{\partial}{\partial d} E_1^{\rm inner}(L_1,L_2,d) \right)\; , \label{pdd}
\eea

\noindent
where we have taken into account that $E_1^{\rm inner}=E_2^{\rm inner}$.

From (\ref{e1in}) and (\ref{pdd}) we see that the only term which contributes to the pressure at the thermodynamic limit is the first term at the right side of (\ref{e1in}); in fact, only the term $m_1=m_2=0$ contributes (that the terms with $m_1\neq 0$ or $m_2\neq 0$, or both, do not contribute can be proved just in the same way that in section \ref{couter} -- see eqs. (\ref{est1})-(\ref{est2})). Then, we obtain
\be
p_C(d)=\frac{1}{16\pi^2}\zeta (4)\frac{\partial}{\partial d}d^{-3} =-\frac{\pi^2}{480}d^{-4}\; ,
\ee

\noindent
which is the same result obtained in the case without the inner Dirichlet plane!

The above result shows that the introduction of additional stresses parallel to the physical plates do not modify the pressure on these plates. Finally, it is easy to see that the above proof generalizes in a trivial way to the case in which the Dirichlet plane is at the outer region, and also to the case in which we have a finite number of Dirichlet planes in the inner region provided these planes are disposed symmetrically with respect to some plane orthogonal to the $z$ axis.

\vspace{1.0cm}

{\bf Note added.} The following related references have been 
brought to our attention: M. Bordag, E. Elizalde, K. Kirsten,
S. Leseduarte, {\it Phys. Rev. D} {\bf 56}, 4896 (1997) for
spherical geometries; and N. F. Svaiter, B. F. Svaiter, 
{\it J. Phys. A} {\bf 25}, 979 (1992) and {\it Phys. 
Rev. D} {\bf 47}, 4581 (1993) for the use of auxiliary cavities
in order to treat surface divergencies.

\end{document}